\documentclass[twocolumn, floatfix, aps, prl, 10pt]{revtex4-2}

\usepackage[utf8]{inputenc}
\usepackage[normalem]{ulem}
\usepackage{graphicx, array, multirow, tabularx, enumitem}
\usepackage{amssymb, braket, mathtools}
\usepackage[dvipsnames]{xcolor}
\usepackage{bm, bbm, dsfont}
\usepackage[urlcolor=blue, colorlinks=true, citecolor=blue,linkcolor=blue]{hyperref}

\newcommand{\vh}{\vec h}

\newcommand{\vq}{\vec q}

\newcommand{\ve}{\vec e}

\newcommand{\vsigma}{\mbox{\boldmath $\sigma$}}

\renewcommand{\vec}[1]{\mathbf{#1}}
\graphicspath{{./}{./figures/}}

\begin{document}

\title{Proximity superconductivity in chiral kagome antiferromagnets}
\author{Adam Yanis Chaou}
\author{Gal Lemut}
\author{Felix von Oppen}
\author{Piet W.\ Brouwer}
\affiliation{Freie Universität Berlin, Dahlem Center for Complex Quantum Systems and Fachbereich Physik, Arnimallee 14, 14195 Berlin}

\begin{abstract}
  Recent experiments on the chiral kagome antiferromagnet Mn$_3$Ge have provided strong evidence of proximity-induced spin-polarized superconductivity. We introduce and explore a minimal model which exhibits a rich phase diagram as a function of chemical potential and spin canting. We find a valley-singlet superconducting phase for chemical potentials and canting consistent with the experimental system. This phase transitions into a Chern insulator at larger canting and gives way to topological superconducting phases with Chern numbers ${\cal C}_{\rm BdG} = \pm 1, \pm 3$ at other chemical potentials. Our results show that proximity-induced superconductivity in kagome antiferromagnets is a promising route towards exotic superconductivity with spin-polarized Cooper pairs, with potential applications in spintronics.
\end{abstract}

\maketitle

\emph{Introduction.—} 
Exchange fields rapidly dephase Cooper pairs formed by electrons of opposite spin, inhibiting conventional superconducting order and suppressing the superconducting proximity effect in ferromagnets. In contrast, triplet superconductivity based on equal-spin Cooper pairs can coexist with magnetic order, because the exchange field polarizes the Cooper pairs, but 
does not dephase them. Superconductors with spin-polarized Cooper pairs can carry dissipationless spin currents, which makes them of interest for spintronic applications \cite{eschrig2015,linder2015}, and require an unconventional superconducting order, such as odd-momentum, odd-frequency, or valley-singlet pairing.

A promising avenue for realizing superconductors with spin-polarized Cooper pairs is via proximity superconductivity in non-collinear magnetic systems. The spatially varying magnetization converts opposite-spin Cooper pairs injected from a conventional spin-singlet superconductor into spin-polarized Cooper pairs, which are robust against the exchange field \cite{kadigrobov2001,volkov2003,houzet2007}. Such proximity-induced spin-polarized superconductivity has been observed in the half-metal CrO$_2$ \cite{keizer2006,anwar2010}, in magnetic multilayers with non-collinear magnetization \cite{khaire2010,anwar2012} or magnetically disordered interfaces \cite{sprungman2010,wang2010}, and in magnetic structures with a helical magnetization profile \cite{sosnin2006,robinson2010}.

\begin{figure*}[t]
	\includegraphics[width=1\textwidth]{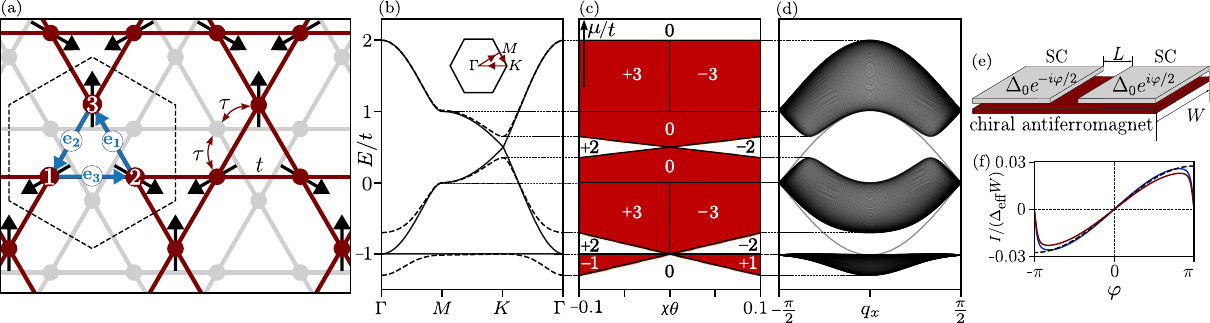}
	\caption{(a) Kagome lattices for an itinerant chiral antiferromagnet (red) and a spin-singlet superconductor (grey). The two layered lattices are displaced in-plane such that the downward-facing triangles of one lattice are on top of upward-facing triangles of the other. The figure displays the unit cell (dashed) with sites $i=1,2,3$ and unit vectors $\ve_{i=1,2,3}$, as well as the amplitudes $t$ for intra-layer hopping in the antiferromagnetic layer and $\tau$ for inter-layer hopping between magnet and superconductor. The spin texture (arrows) corresponds to $\chi = 1$, $\phi = \pi/2$. (b--d) Band structure of the itinerant chiral antiferromagnet in the fully-polarized limit of Eq.\ (\ref{eq:H_AFM}), phase diagram depicting the Chern number ${\cal C}_{\rm BdG}$ for the full BdG Hamiltonian ${\cal H}'$ and the presence of superconducting correlations (red) as a function of the chemical potential $\mu$ and the canting angle $\theta$, and the spectrum of the normal-state Hamiltonian (\ref{eq:H_AFM}) for a strip extended in the $x$ direction with canting angle $\theta = 0.1$. Chiral edge states connecting the bulk bands reflect the topology of the normal-state band structure. (e) Josephson junction, consisting of an itinerant chiral antiferromagnet (red), proximity coupled to two superconductors with phase difference $\varphi$ (grey). (f) Current-phase relation for chemical potential $\tilde \mu = -0.1 t$ (blue), $0.1 t$ (red) near filling fraction $\nu = 2$. The analytical prediction (dashed) for the low-energy model (\ref{eq:valley_singlet}) is compared with a numerical calculation of the full lattice model (solid) for $\Delta_{\rm eff} = 0.005 t$.}
	\label{fig:Panel}
\end{figure*}

Recently, the observation of a sizable Josephson effect in the non-collinear antiferromagnet Mn$_3$Ge over distances $\gtrsim 200$ nm was interpreted as a signature of spin-polarized proximity superconductivity \cite{jeon2021,jeon2023}. In this material, the Mn atoms form two displaced kagome layers and order magnetically into a chiral $120^\circ$ antiferromagnet, see Fig.\ \ref{fig:Panel}a \cite{nayak2016,kiyohara2016,yang2017,zhang2017,soh2020}. The combination of microscopic non-collinearity of the magnetic order with zero net magnetization enables efficient conversion of spin-singlet Cooper pairs into long-range spin-polarized supercurrents. 

Here, we propose a minimal model consisting of a two-dimensional itinerant chiral non-collinear antiferromagnet coupled to an s-wave superconductor that captures the observed behavior. We consider the fully-polarized limit that the exchange field is larger than the bandwidth \cite{ohgushi2000spin}, so that any forms of opposite-spin proximity superconductivity are manifestly ruled out \cite{eschrig2003,galaktionov2008,eschrig2008,beri2009}. We find that, if inversion symmetry is broken by the coupling to the superconductor, the non-collinear nature of antiferromagnetic order enables the induction of a fully gapped superconducting phase with (spatially textured) spin-polarized Cooper pairs. 
We explore the phase diagram of the model as a function of Fermi energy and out-of-plane canting angle $\theta$. In the regime relevant to Mn$_3$Ge (Fermi pockets around the $K$ and $K'$ points, small $\theta$), we find a valley-singlet superconducting phase. This phase transitions into a Chern insulator at larger canting angles. We also identify a rich set of topological superconducting and insulating phases for other Fermi energies.

Below, we first describe the appearance of spin-polarized proximity superconductivity in a monolayer of the kagome antiferromagnet. For a single-layer antiferromagnet, the condition that inversion symmetry must be broken requires that the coupling between the kagome magnet and the superconductor be spatially non-uniform. This is naturally realized if the superconductor has the same lattice structure as the kagome magnet, but with a spatial shift, such that the downward-facing triangles of one lattice are on top of the upward-facing triangle of the other, see Fig.\ \ref{fig:Panel}a. We also consider a bilayer kagome antiferromagnet, for which spatially uniform coupling to a superconductor already breaks inversion symmetry and show that spin-polarized superconductivity can be induced in this system as well.

\emph{Chiral antiferromagnet.—} We describe the itinerant chiral antiferromagnet as a two-dimensional kagome lattice with nearest-neighbor hopping and an on-site exchange field that imposes a chiral spin texture \cite{chen2014}. The Bloch Hamiltonian reads
\begin{equation}
  H_{ij}(\vq) = -\vsigma \cdot \vh_j \delta_{ij} + \sigma_0 T_{ij}(\vq),
  \label{eq:H_Bloch}
\end{equation}
where the vector of Pauli matrices $\vsigma$ as well as the unit matrix $\sigma_0$ act in spin space, and the indices $i,j=1,2,3$ refer to the three sites in the unit cell of the kagome lattice, see Fig.\ \ref{fig:Panel}a. The first term in Eq.\ (\ref{eq:H_Bloch}) represents the interaction with the exchange field
\begin{equation}
  \vh_{j} = h (\ve_z \sin \theta + \ve_x \cos\theta \cos \phi_j +
  \ve_y \cos \theta \sin \phi_j),
  \label{eq:hh}
\end{equation}
for which the azimuthal angle
$\phi_j = \phi + j \omega$
differs by $\omega = \frac{2 \pi}{3} \chi$ between the sites in the unit cell and $\chi = \pm 1$ is the vector chirality of the spin texture. The out-of-plane canting angle $\theta$ is the same for all spins. The second term in Eq.\ (\ref{eq:H_Bloch}) accounts for nearest-neighbor hopping on the kagome lattice with amplitude $t$,
\begin{equation}
    T(\vec{q}) = -2t \begin{pmatrix}
    0 & \cos q_3 & \cos q_2 \\
    \cos q_3  & 0 & \cos q_1 \\
    \cos q_2 & \cos q_1 & 0 \end{pmatrix},
    \label{eq:Hspinful}
\end{equation}
where we assume unit bond length and $q_i = \vq \cdot \ve_i$ with $\ve_{i=1,2,3}$ the vectors connecting neighboring lattice sites, see Fig.\ \ref{fig:Panel}a.

The chiral antiferromagnet is tunnel-coupled to an s-wave superconductor, which we model as a second kagome layer but with superconducting order (order parameter $\Delta_0 = |\Delta_0| e^{i \varphi}$) instead of an exchange field. The kagome lattices of the superconductor and the antiferromagnet are displaced in-plane with respect to each other in the same manner as the individual kagome layers of Mn$_3$Ge, {\em i.e.}, the downward-facing triangles of the superconductor  are on top of the upward-facing triangles of the antiferromagnet, see Fig.\ \ref{fig:Panel}a. One may view the superconducting kagome layer as a boundary layer of the chiral magnet that has lost its magnetic order because of proximity to the superconductor, so that it admits conventional spin-singlet proximity superconductivity. The in-plane displacement between the magnetic and superconducting kagome layers ensures that in-plane inversion symmetry is broken, which is a necessary condition for spin-polarized superconductivity if the bands are non-degenerate. The need for a spatially non-uniform coupling to the superconductor is special to a monolayer antiferromagnet. For multilayer magnets, inversion symmetry is automatically broken upon coupling the topmost layer to the superconductor. In this case we also find spin-polarized proximity superconductivity for a uniform coupling to the superconductor, as discussed further below.

Tunneling between the superconducting and magnetic layers is dominated by the tunneling amplitude $\tau$ between the closest pairs of sites in the two lattices, as indicated in Fig.\ \ref{fig:Panel}a. Since the superconductor is gapped, its degrees of freedom may be integrated out, and we arrive at an effective $12 \times 12$ Bogoliubov-de Gennes (BdG) Hamiltonian ${\cal H}(\vq)$ describing the proximity superconductivity in the chiral antiferromagnet. To leading order in the interlayer tunneling amplitudes and using the zero-bandwidth approximation for the superconductor one has
\begin{equation}
  {\cal H}(\vq) =
  \begin{pmatrix} H(\vq) - \mu & \Delta(\vq) \\ -\Delta(-\vq)^* & -H^*(-\vq) + \mu \end{pmatrix},
  \label{eq:HBDG}
\end{equation}
where $\mu$ is the chemical potential and
\begin{equation}
  \label{eq:Delta_eff}
  \Delta(\vq) =
  \Delta_{\rm eff}
  \sigma_y
  \begin{pmatrix}
    2 & e^{i q_3} & e^{-i q_2} \\ e^{-i q_3} & 2 & e^{i q_1} \\
    e^{i q_2} & e^{-i q_1} & 2
  \end{pmatrix}
\end{equation}
in terms of the effective, proximity-induced superconducting order parameter $\Delta_{\rm eff} = |\tau|^2/\Delta_0^* = |\Delta_{\rm eff}| e^{i \varphi}$ in the chiral antiferromagnet. 

{\em Fully-polarized limit.—}
To simplify the discussion, we focus on the regime $h \gg t$ in which the exchange field is the dominant energy scale. For $h \gg t$, the normal-state Hamiltonian $H(\vq)$ of Eq.\ (\ref{eq:H_Bloch}) has three bands each around the energies $\pm h$, with electron spins aligned parallel and antiparallel to the exchange field. Projecting onto the lower three bands \cite{ohgushi2000spin} and shifting all normal-state energies by $h$, we obtain the $6 \times 6$ BdG Hamiltonian
\begin{equation}
  {\cal H}'(\vq) = \begin{pmatrix} H'(\vq) - \mu & \Delta'(\vq) \\ -\Delta'(-\vq)^* & -H'^*(-\vq) + \mu \end{pmatrix}.
  \label{eq:HM}
\end{equation}
Here
\begin{equation}
    H'(\vec{q}) = -2 \begin{pmatrix}
    0            & t' \cos q_3   & t'^* \cos q_2   \\
    t'^* \cos q_3 & 0            & t' \cos q_1  \\
    t' \cos q_2   &  t'^* \cos q_1  & 0    
    \end{pmatrix},
    \label{eq:H_AFM}
\end{equation}
where the complex hopping amplitude
\begin{align}
  t' = t e^{i \omega/2} [\cos (\omega/2) - i \sin (\omega/2) \sin \theta]
\end{align}
includes the overlap between two spin-$1/2$ states with azimuthal angles differing by $\omega$. (Note that $|t'| = t/2$ for $\theta = 0$.)
The projected effective order parameter is
\begin{align}
  \label{eq:Delta_eff_prime}
  \Delta'(\vq) =&\,
  \Delta'_{\rm eff}
  \begin{pmatrix}
    0 & e^{i q_3} & - e^{-i q_2 + i \omega} \\ - e^{-i q_3} & 0 & e^{i q_1 - i \omega} \\
    e^{i q_2 + i \omega} & -e^{-i q_1 - i \omega} & 0
  \end{pmatrix},  
\end{align}
with $\Delta'_{\rm eff} = \Delta_{\rm eff} e^{- i \phi} \cos \theta \sin \omega$. The diagonal elements in Eq.\ (\ref{eq:Delta_eff_prime}) vanish because the exchange field prohibits spin-singlet Cooper pairs on the same lattice site. The off-diagonal elements are nonzero, because a spin-singlet Cooper pair in the superconductor has a finite matrix element for hopping to a pair of neighboring lattice sites as these have different spin directions. 

In the coplanar limit, $\theta = 0$, the normal-state Hamiltonian $H'(\vq)$ describes a kagome lattice with a flux $\pi$ through the triangles. It has a gapless band structure that is identical to that of a kagome lattice without flux, up to an overall minus sign, see Figs.\ \ref{fig:Panel}b, d. \cite{ohgushi2000spin}. There is a flat band at energy $\varepsilon = -t$ and a pair of dispersive bands with energies between $-t$ and $2 t$, which touch the flat band at $\Gamma$ and each other at $K$ and $K'$. Canting the spins out of the plane lifts the degeneracies. After canting, the system hosts three separate bands with Chern numbers $\mathcal{C}=\pm 1, 0, \mp 1$ (from top to bottom), where the sign of the Chern numbers is determined by the sign of $\chi \theta$ (upper signs for $\chi \theta > 0$).

{\em Phase diagram.—}
If the chemical potential $\mu$ lies in one of the bands of $H'(\vq)$, the proximity coupling to the superconductor induces superconducting correlations. Figure \ref{fig:Panel}c shows the Chern number ${\cal C}_{\rm BdG}$ of the proximity-induced superconducting phase as a function of the chemical potential $\mu$ and the canting angle $\theta$. In the vicinity of filling fractions $\nu = 0$, $1$, and $3$ (corresponding to $\mu$ near $-t$ or $2 t$), we find gapped chiral superconducting phases with odd Chern numbers ${\cal C}_{\rm BdG}$ for $\theta \neq 0$. The chirality of the pairing is set by the sign of the canting angle $\theta$ and the superconductor becomes nodal at the topological phase transition at $\theta = 0$. Near filling fraction $\nu = 2$, the normal state has Fermi pockets near the $K$ and $K'$ points. Here we find a fully gapped, but topologically trivial valley-singlet phase. If the chemical potential $\mu$ lies inside one of the two gaps of the normal-state Hamiltonian $H'(\vq)$, which requires $\theta \neq 0$, the system is a non-superconducting Chern insulator with Chern number $|{\cal C}| = 1$. (Correspondingly, the full BdG Hamiltonian has Chern number $|{\cal C}_{\rm  BdG}| = 2$, because of the Hilbert-space doubling.)

{\em Valley-singlet phase near $\nu = 2$.—} 
To analyze the proximity superconductivity in more detail, we now focus on the vicinity of the filling fraction $\nu = 2$. The Fermi pockets around the $K$ and $K'$ points correspond to the position of the Fermi pockets for the chiral antiferromagnet Mn$_3$Ge of Refs.\ \cite{jeon2021,jeon2023}. We consider the limit of small canting $|\theta| \ll 1$, for which $\nu=2$ corresponds to the chemical potential $\mu = t/2$. Writing $\mu = t/2 + \tilde{\mu}$, we obtain the low-energy $4 \times 4$ BdG Hamiltonian \begin{equation}
  \tilde {\cal H}(\tilde{\vq}) = \begin{pmatrix} \tilde H_+(\tilde{\vq}) - \tilde{\mu}& \tilde \Delta \\
  \tilde \Delta^{\dagger} & -\tilde H_-(-\tilde{\vq})^* + \tilde{\mu} \end{pmatrix},
  \label{eq:HK}
\end{equation}
describing electron-like states at $K$ ($+$) and hole-like states at $K'$ ($-$). Here $\tilde{\vq}$ is the reciprocal-space distance from the $K$ point and, to lowest order in $\tilde{\vq}$, $\theta$ and $\Delta_{\rm eff}'$,
\begin{align} 
  \label{eq:eff_Dirac}  
  \tilde H_{\pm}(\tilde{\vq}) =&\, 
  \begin{pmatrix}
    \varepsilon_{\theta} & \mp v \tilde{q} e^{- i \alpha} \\
  \mp v \tilde{q} e^{i \alpha} & - \varepsilon_{\theta}
  \end{pmatrix}, \\
  \label{eq:eff_DiracDelta}
  \tilde \Delta =&\, 
  \begin{pmatrix} 0 & 0 \\ -i \Delta_{\rm eff}' \sqrt{3} & 0 \end{pmatrix},
\end{align}
where we abbreviated $\varepsilon_{\theta} = ({3t}/{2}) \theta \chi$, $v = ({t}/{2}) \sqrt{3}$, and $\tilde{q} e^{i \alpha} = \tilde{q}_x + i \tilde{q}_y$.
The full BdG Hamiltonian also contains a block describing electron-like states at $K'$ and hole-like states at $K$, which is the particle-hole conjugate of $\tilde {\cal H}(\tilde \vq)$.

The system is in a fully gapped valley-singlet superconducting phase if $|\tilde{\mu}| > |\varepsilon_{\theta}|$. In the limit $|\Delta_{\rm eff}'| \ll |\varepsilon_{\theta}| \ll |\tilde{\mu}|$, this phase is described by the effective BdG Hamiltonian
\begin{equation}
  \tilde {\cal H}_{\rm eff}(\tilde{\vq}) = \begin{pmatrix} 
    \pm v (\tilde{q} - \tilde{q}_{\rm F}) &  
  \pm \frac{i}{2} \Delta_{\rm eff}' e^{\mp i s_{\theta} \alpha} \sqrt{3}  \\
  \mp \frac{i}{2} \Delta_{\rm eff}'^* e^{\pm i s_{\theta} \alpha} \sqrt{3}  & \mp v (\tilde{q} - \tilde{q}_{\rm F})
  \end{pmatrix},
  \label{eq:valley_singlet}
\end{equation}  
where $\tilde \mu = \pm \tilde{q}_{\rm F} v$ and $s_{\theta} = \mbox{sgn}\, \varepsilon_{\theta}$. This phase necessarily has Chern number $\mathcal{C}_{\rm BdG} = 0$, as it connects without gap closing to the coplanar limit $\theta = 0$, which has an effective time-reversal symmetry, see the phase diagram of Fig.\ \ref{fig:Panel}c.

The transition between the insulating and superconducting phases for $|\tilde \mu| < |\varepsilon_{\theta}|$ and $|\tilde \mu| > |\varepsilon_{\theta}|$ can be captured by expanding around $\tilde{\mu} = \pm \varepsilon_{\theta} + d\tilde \mu$ for small $|\tilde{\vq}|$, $|d\mu| \ll |\varepsilon_{\theta}|$. Here, one finds the effective Hamiltonian
\begin{align}
  \label{eq:valley_transition}
  \tilde H_{\rm eff}(\tilde \vq) =&\,
  \begin{pmatrix}
    \pm \tfrac{v^2 \tilde q^2}{2 \varepsilon_{\theta}} - d\tilde \mu &
      \pm  i \tfrac{v \tilde q}{2 \varepsilon_{\theta}}
      \Delta_{\rm eff}' e^{\mp i \alpha} \sqrt{3} \\
      \mp i \tfrac{v \tilde q}{2 \varepsilon_{\theta}}
      \Delta_{\rm eff}'^* e^{\pm i \alpha} \sqrt{3} &
      d \tilde \mu \mp  \tfrac{v^2 \tilde q^2}{2 \varepsilon_{\theta}}
  \end{pmatrix},
\end{align}
This model undergoes a topological phase transition as $d\tilde \mu$ changes sign. The Chern number changes by one across the transition, with the same contribution from the opposite valley, so that $|\delta {\cal C}_{\rm BdG}| = 2$.

\emph{Josephson junction.—}
We next consider a Josephson junction, consisting of a two-dimensional chiral antiferromagnet, which is tunnel-coupled to two different superconductors for $y > L/2$ and $y < -L/2$, with a phase difference $\varphi$ between them, see Fig.\ \ref{fig:Panel}e. The lattice Hamiltonian of such a junction is described by (the Fourier transform of) Eq.\ (\ref{eq:HBDG}), but with $\Delta_{\rm eff}(y) = |\Delta_{\rm eff}| e^{i \varphi\, {\rm sign}(y)/2} \Theta(|y| - L/2)$. We again focus on the valley-singlet superconductor in the vicinity of filling fraction $\nu = 2$, where we can use the effective low-energy Hamiltonian (\ref{eq:valley_singlet}). Since there is translation symmetry in the $x$ direction, each transverse mode (parameterized by the momentum $-q_{\rm F} < \tilde{q}_x < q_{\rm F}$) contributes independently to the Josephson current \cite{beenakker1991d}. Using the well-known result for the Josephson current in a ballistic one-dimensional channel in the short-junction limit $\hbar L v_{\rm F} \ll |\Delta_{\rm eff}'|$ \cite{haberkorn1978,zaitsev1984,arnold1985,furusaki1990} and with superconducting gap $(1/2) |\Delta_{\rm eff}'| \sqrt{3} = (3/4) |\Delta_{\rm eff}|$, one thus finds
\begin{equation}
  I(\varphi) = \frac{e}{\hbar} \frac{3 W \tilde q_{\rm F}}{2 \pi} |\Delta_{\rm eff}| \sin \frac{\varphi}{2},
  \ \ -\pi < \varphi < \pi,
  \label{eq:JJ}
\end{equation}
where $W \gg L$ is the junction width.

We compare Eq.\ (\ref{eq:JJ}) to numerical simulations of a short Josephson junction for the full lattice model (\ref{eq:HBDG}) using the kwant package \cite{groth2014}. Figure \ref{fig:Panel}f shows the current-phase relationship in the valley-singlet phase with filling just below and above $\nu = 2$. (If the filling is precisely equal to $\nu = 2$ the system is in an insulating phase and the supercurrent vanishes.) The differences between the analytical prediction of Eq.\ (\ref{eq:JJ}) and the numerical calculations result from the finite values of $|\tilde \mu|/t$ and $|\Delta_{\rm eff}'|/t$ used in the simulations, whereas these quantities need to be $\ll 1$ for the effective low-energy theory of Eq.\ (\ref{eq:valley_singlet}), which is used to derive Eq.\ (\ref{eq:JJ}).

\begin{figure}[t]
    \includegraphics[width=1\columnwidth]{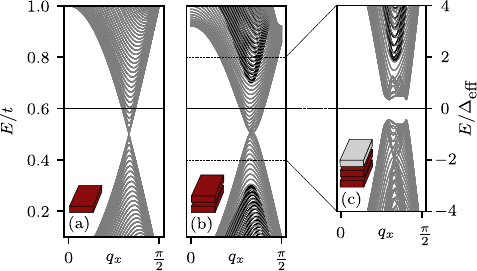}
	\caption{(a) Spectrum of the normal-state Hamiltonian $H'$ of Eq.\ (\ref{eq:HM}) for a strip extended in the $x$ direction and canting angle $\theta = 0$. (b) Spectrum of the normal-state bilayer of two chiral antiferromagnets on the kagome lattice with interlayer coupling $\tau = 0.2 t$. (c) Spectrum of the BdG Hamiltonian ${\cal H}'(\vq)$ of a bilayer proximitised with a spatially uniform superconductor with interlayer coupling $\tau = 0.2 t$ and $\Delta_{\rm eff} = 0.05 t$. The chemical potential $\mu = 0.6 t$ is indicated by the horizontal solid line in (a) and (b). 
	\label{fig:Bilayer}}
\end{figure}

{\em Induced superconductivity in a bilayer magnet.—} In a multilayer chiral antiferromagnet, proximity coupling to the superconductor breaks inversion symmetry even if the coupling is spatially uniform. We illustrate this by considering a bilayer magnet, with one layer coupled to the superconductor and the magnetic kagome layers shifted in-plane as in Mn$_3$Ge. We refer to the appendix for lattice models and an effective low-energy theory near filling fraction $\nu = 2+2$. Spectra for a bilayer in a strip geometry displayed in Fig.\ \ref{fig:Bilayer} show that coupling to a spatially uniform superconductor opens a full spectral gap. Since the bilayer is in the fully-polarized limit, opposite-spin pairing is ruled out, so that the gap-opening is indicative of spin-polarized superconductivity.

\emph{Conclusion.—} 
We presented a minimal microscopic model for spin-polarized proximity superconductivity in an itinerant chiral antiferromagnet on the kagome lattice. We considered the limit of large exchange fields, which ensures that all forms of opposite-spin proximity superconductivity are manifestly ruled out. In the experimentally relevant situation with small Fermi pockets near the $K$ and $K'$ points, the proximity-induced superconducting phase is of valley-singlet type. This is a topologically trivial phase, which persists for canting angle $\theta = 0$. For other filling fractions we find topological phases with Chern numbers $|{\cal C}_{\rm BdG}| =1$, $3$, which require a nonzero canting angle $\theta$, as well as nodal superconducting phases at the topological phase transition at $\theta = 0$.

References \cite{jeon2021,jeon2023} attribute the presence of proximity-induced spin-polarized superconductivity in Mn$_3$Ge to the presence of a large Berry curvature imposed by the non-collinear texture. The Berry curvature acts as an effective magnetic field in reciprocal space, which converts spin-singlet Cooper pairs injected by the superconductor into spin-polarized pairs. Our model shows that spin-polarized proximity superconductivity can exist without Berry curvature. The Berry curvature of the normal-state Hamiltonian $H(\vq)$ vanishes everywhere in reciprocal space in the coplanar limit $\theta = 0$. Nevertheless, we find a bona fide spin-polarized superconducting phase for $\theta = 0$ near $\nu = 2$, see Eq.\ (\ref{eq:valley_singlet}) and Fig.\ \ref{fig:Panel}c. Conversely, breaking inversion symmetry ({\em e.g.}, by having different hopping amplitudes for bonds in upward and downward-facing triangles) results in Berry curvature locally in reciprocal space even in the collinear limit $|\theta| = \pi/2$, but spin-polarized induced superconductivity is ruled out in this case. Both observations point to non-collinearity of spins, and not Berry curvature, as the sole driving force of spin-polarized proximity superconductivity.

Chiral antiferromagnets have been proposed as a medium for efficient spintronic information storage and processing \cite{baltz2018}. Switching between magnetic ground states as well as readout of the magnetic state by electric currents have been demonstrated experimentally \cite{tsai2020,takeuchi2021,pal2022,chen2023}. Establishing chiral antiferromagnets as a platform for spin-polarized superconductivity paves the way towards replacing dissipative normal currents by dissipationless supercurrents in these spintronic applications.

{\em Acknowledgements.—} 
We would like to thank Banabir Pal, Stuart Parkin, Christoph Strunk, and Georg Woltersdorf for helpful discussions and Isidora Araya Day and Anton Akhmerov for assistance with Pymablock \cite{pymablock_code, pymablock_paper}. This work was supported by the Deutsche Forschungsgemeinschaft (DFG, German Research Foundation) - Project Number 277101999 - CRC TR 183 (projects A02, A03, and C03) and stimulated by the Cluster of Excellence Exc 3112 Center for Chiral Electronics. When this manuscript was close to completion, two articles appeared, which also considered the Josephson effect and spin-polarized superconductivity in two-dimensional chiral antiferromagnets \cite{hou2025,zhang2025}. 

\appendix

\section{Model symmetries}
\label{app:symm}

For a coplanar spin texture, $\theta = 0$, the lattice model (\ref{eq:Hspinful}) has three-fold rotation ${\cal R}$, inversion ${\cal I}$, and mirror ${\cal M}$ symmetries, as well as an effective spinless time-reversal symmetry ${\cal T}$. If $\theta \neq 0$, the ${\cal M}$ and ${\cal T}$ symmetries are broken, but the product ${\cal MT}$ remains preserved. Below, we discuss these symmetries and their representations.

We first consider the action of the spatial symmetry operations ${\cal R}$, ${\cal I}$, and ${\cal M}$ on the unit vectors $\ve_i$ and the exchange field $\vh$.
The unit vectors $\ve_i$ satisfy
\begin{equation}
  R \ve_1 = \ve_2,\ \ R \ve_2 = \ve_3,\ \ R \ve_3 = \ve_1,
\end{equation}
with $R$ a rotation by $2 \pi/3$ around the center of the unit cell,
\begin{equation}
  M \ve_1 = -\ve_2,\ \ M \ve_2 = -\ve_1,\ \
  M \ve_3 = - \ve_3,
\end{equation}
with $M$ the mirror reflection about the vertical in Fig.\ \ref{fig:Panel}a, and 
\begin{equation}
  I \ve_1 = - \ve_1,\ \ I \ve_2 = - \ve_2,\ \ I \ve_3 = - \ve_3,
\end{equation}
with $I$ the inversion about one of the sites of the kagome lattice. The exchange field satisfies the symmetry relations
\begin{equation}
  \vsigma \cdot \vh_{\theta,\phi - \omega} = e^{i \omega \sigma_3/2}
  (\vsigma \cdot \vh_{\theta,\phi}) e^{-i \omega \sigma_3/2}
\end{equation}
and
\begin{align}
  \vsigma \cdot \vh_{\theta,\phi \pm \omega} =&\, 
  \sigma_{\phi} (\vsigma \cdot \vh_{-\theta,\phi \mp \omega}) \sigma_{\phi},
\end{align}
with
\begin{equation}
  \sigma_{\phi} = \sigma_1 \cos \phi + \sigma_2 \sin \phi.
\end{equation}

One then finds that the spinful lattice model of Eq.\ (\ref{eq:Hspinful}) satisfies a threefold rotation symmetry, \begin{equation}
  H(\vq) = {\cal R}^{-1} H(R \vq) {\cal R},
\end{equation}
with
\begin{equation}
  {\cal R} = e^{i \omega \sigma_3/2} \begin{pmatrix} 0 & 0 & 1 \\ 1 & 0 & 0 \\ 0 & 1 & 0 \end{pmatrix},
\end{equation}
as well as inversion symmetry
\begin{equation}
  H(\vq) = {\cal I}^{-1} H(-\vq) {\cal I}
  \label{eq:Hinv}
\end{equation}
with ${\cal I} = \openone$.
For $\theta = 0$, the spinful lattice model of Eq.\ (\ref{eq:Hspinful}) also satisfies a mirror symmetry combined with a $\pi$ rotation of the spins around the $z$ axis,
\begin{equation}
  H(\vq) = {\cal M}^{-1} H(M \vq) {\cal M},
\end{equation}
with
\begin{equation}
  {\cal M} = \begin{pmatrix} 0 & 1 & 0 \\ 1 & 0 & 0 \\ 0 & 0 & 1 \end{pmatrix}
  \sigma_{\phi},
\end{equation}
as well as a ``spinless time-reversal'' symmetry
\begin{equation}
  H(\vq) = {\cal T}^{-1} H(-\vq) {\cal T},
\end{equation}
with
\begin{equation}
  {\cal T} = \sigma_1 K,
\end{equation}
with $K$ complex conjugation. If $\theta \neq 0$, ${\cal M}$ and ${\cal T}$ are no longer good individual symmetries, but the product ${\cal M} {\cal T}$ remains a good symmetry,
\begin{equation}
  H(\vq) = ({\cal M} {\cal T})^{-1} H(-M\vq) ({\cal M}{\cal T}).
\end{equation}

\section{Fully-polarized limit}

In the fully-polarized limit $h \gg t$, the lowest three bands have the spinor 
\begin{equation}
  \ket{j} = \begin{pmatrix} \cos(\pi/4-\theta/2) \\ e^{i \phi_j} \sin(\pi/4-\theta/2) \end{pmatrix}
\end{equation}
at site $j$ in the unit cell, $j=1,2,3$. This gives the normal-state Hamiltonian
\begin{equation}
  H'(\vq) = -2 t \begin{pmatrix} 
  0 & \bra{1}\! 2\rangle \cos q_3 & \bra{1}\! 3\rangle \cos q_2 \\ 
  \bra{2}\! 1\rangle \cos q_3 & 0 & \bra{2}\! 3\rangle \cos q_1 \\
  \bra{3}\! 1\rangle \cos q_2 & \bra{3}\! 2\rangle \cos q_1 & 0
  \end{pmatrix}.
\end{equation}
The overlap matrix elements are
\begin{align}
  \bra{1}\! 2\rangle =&\,
  \bra{2}\! 3\rangle = 
  \bra{3}\! 1\rangle \nonumber \\ =&\,
  e^{i \omega/2}(\cos(\omega/2) - i \sin(\omega/2) \sin \theta),
\end{align}
so that Eq.\ (\ref{eq:H_AFM}) follows immediately. Similarly, we find that
\begin{align}
  \lefteqn{\Delta'(\vq) = \Delta_{\rm eff}} \\ \nonumber &\, \mbox{} \times
  \begin{pmatrix}
  2 \bra{1}\! \sigma_y\! \ket{1}^* &
  e^{i q_3} \bra{1}\! \sigma_y\! \ket{2}^* & 
  e^{-i q_2} \bra{1}\! \sigma_y\! \ket{3}^* \\
  e^{-i q_3} \bra{2}\! \sigma_y\! \ket{1}^* & 
  2 \bra{2}\! \sigma_y\! \ket{2}^* &
  e^{i q_1} \bra{2}\! \sigma_y\! \ket{3}^* \\
  e^{i q_2} \bra{3}\! \sigma_y\! \ket{1}^* & 
  e^{-i q_1} \bra{3}\! \sigma_y\! \ket{2}^* &
  2 \bra{3}\! \sigma_y\! \ket{3}^*
  \end{pmatrix}.
\end{align}
Using the matrix elements
\begin{align}
  \bra{1}\! \sigma_y\! \ket{1}^* = &\,
  \bra{2}\! \sigma_y\! \ket{2}^* \nonumber \\ =&\,
  \bra{3}\! \sigma_y\! \ket{3}^* \nonumber \\ =&\, 0, \nonumber \\ 
  \bra{1}\! \sigma_y\! \ket{2}^* = &\, - \bra{2}\! \sigma_y\! \ket{1}^* \nonumber \\ \,
  =&\, e^{-i \phi} \cos \theta \sin \omega \nonumber \\
  \bra{2}\! \sigma_y\! \ket{3}^* = &\, - \bra{3}\! \sigma_y\! \ket{2}^* \nonumber \\\,
  =&\, e^{-i \phi - i \omega} \cos \theta \sin \omega \nonumber \\
  \bra{3}\! \sigma_y\! \ket{1}^* = &\, - \bra{1}\! \sigma_y\! \ket{3}^* \nonumber \\\,
  =&\, e^{-i \phi + i \omega} \cos \theta \sin \omega,
\end{align}
one recovers Eq.\ (\ref{eq:Delta_eff_prime}) of the main text.\bigskip

\section{Broken inversion symmetry as a prerequisite for spin-polarized superconductivity}

While the normal-state Hamiltonian $H(\vq)$ of Eqs.\ (\ref{eq:H_Bloch}) and (\ref{eq:H_AFM}) of the main text is inversion symmetric, $H(\vq) = H(-\vq)$, see Eq.\ (\ref{eq:Hinv}), the effective superconducting order parameter $\Delta(\vq)$ of Eqs.\ (\ref{eq:Delta_eff}) and (\ref{eq:Delta_eff_prime}) breaks inversion symmetry, because the proximity coupling exists for the upward-facing triangles in the kagome lattice only, see Fig.\ \ref{fig:Panel}a. Breaking of inversion symmetry is a prerequisite for inducing a superconducting gap in a system with spin-split bands: Combined with particle-hole symmetry, $\Delta(\vq) = -\Delta(-\vq)^{\rm T}$, inversion symmetry would impose that $\Delta(\vq) = -\Delta(\vq)^{\rm T}$ is an antisymmetric matrix in the Bloch basis, a property that is preserved upon transforming to the eigenbasis of $H(\vq)$, since $H(\vq)$ is inversion symmetric. It is therefore impossible to induce a superconducting gap in a non-degenerate band unless inversion symmetry is broken.

\section{Effective low-energy theory around filling \texorpdfstring{$\nu = 2$}{}}

We have used the geometric positions of the three sites in the unit cell to define the Bloch Hamiltonian (\ref{eq:H_Bloch}). As a consequence, intra-unit-cell hoppings are also $\vq$-dependent and ${\cal H}(\vq)$ is not periodic for shifts of $\vq$ by a reciprocal lattice vector.

For definiteness, we choose the unit vectors $\ve_i$, $i=1,2,3$, as
\begin{align}
  \label{eq:exy}
    \ve_1 =&\, - \tfrac{1}{2} \ve_x + \tfrac{1}{2} \ve_y \sqrt{3}, \nonumber \\
    \ve_2 =&\, - \tfrac{1}{2} \ve_x - \tfrac{1}{2} \ve_y \sqrt{3}, \nonumber \\
    \ve_3 =&\, \ve_x.
\end{align}
When constructing the effective low-energy theory for filling fraction $\nu \approx 2$, we take the $K$ and $K'$ points at $\vq_{K} = \tfrac{2 \pi}{3} \ve_x$ and $\vq_{K'} = - \tfrac{2 \pi}{3} \ve_x$. For $\vq$ in the vicinity of the $K$ point, we then have $\vq = \tfrac{2 \pi}{3} \ve_x + \tilde \vq$, so that 
\begin{align}
  \label{eq:qKK}
  q_1 =&\, - \tfrac{\pi}{3} - \tfrac{1}{2} \tilde q_x + \tfrac{1}{2} \tilde q_y \sqrt{3}, \nonumber \\ 
  q_2 =&\, -\tfrac{\pi}{3} - \tfrac{1}{2} \tilde q_x - \tfrac{1}{2} \tilde q_y \sqrt{3}, \nonumber \\ 
  q_3 =&\, \tfrac{2 \pi}{3} + \tilde q_x.
\end{align}

The eigenkets of $H'(\vq_{K}) = H(\vq_{K'})$ for $\theta = 0$ at eigenvalue $t'$ are denoted $|\pm\rangle$, with 
\begin{align}
  \label{eq:Keigenkets}
  |+\rangle = \frac{1}{\sqrt{3}} \begin{pmatrix} - 1 \\ -1 \\ 1 \end{pmatrix},\ \
  |-\rangle = \frac{1}{\sqrt{3}} \begin{pmatrix} e^{i \pi \chi/3} \\ e^{-i \pi \chi/3} \\ 1 \end{pmatrix}.
\end{align}
We obtain the effective low-energy Hamiltonian upon restricting the BdG Hamiltonian ${\cal H}'(\vq)$ to the states spanned by the (ordered) basis $\{ |+\rangle,|-\rangle,|+\rangle^*,|-\rangle^* \}$ for $\chi = 1$ and $\{ |-\rangle,|+\rangle,|-\rangle^*,|+\rangle^* \}$ for $\chi = -1$, expanding the diagonal blocks to first order in $\tilde \vq$ and $\theta$, while setting $\tilde \vq$ to zero in the off-diagonal blocks proportional to $\Delta_{\rm eff}'$. (Complex conjugates of the basis kets refer to the hole sector.) This gives the effective low-energy Hamiltonian of the form (\ref{eq:HK}) with
\begin{align}
  \label{eq:eff_DiracDelta_app}
  \tilde H_{\pm}(\tilde{\vq}) =&\,
  \frac{t}{2}
  \begin{pmatrix} 3 \theta \chi & \mp (\tilde q_x - i \tilde q_y) \sqrt{3} \\
    \mp (\tilde q_x + i \tilde q_y) \sqrt{3} & - 3 \theta \chi \end{pmatrix}, 
    \nonumber  \\
    \tilde \Delta(\tilde{\vq}) =&\,
  \begin{pmatrix} 0 & 0 \\ -i \Delta_{\rm eff}' \sqrt{3} & 0 \end{pmatrix}.
\end{align}
With the definitions $\varepsilon_{\theta} = (3/2) t \theta \chi$, $v = (1/2) t \sqrt{3}$, and $\tilde \vq = \tilde q e^{i \alpha}$ one may rewrite these equations as Eqs.\ (\ref{eq:eff_Dirac}) and (\ref{eq:eff_DiracDelta}) of the main text.

We now perform a further projection to obtain effective low-energy theories near the transition at $|\tilde \mu| \approx |\varepsilon_{\theta}|$ and well inside the proximity-induced superconducting phase for $|\tilde \mu| \gg |\varepsilon_{\theta}|$. To this end, we label the basis states in the $2 \times 2$ matrix notation of Eq.\ (\ref{eq:eff_DiracDelta_app}) as $|1\rangle$ and $|2\rangle$. In order to obtain the low-energy theory around $\tilde \mu = \varepsilon_{\theta}$ and $\tilde \mu = - \varepsilon_{\theta}$, we project onto the basis vectors $\{ |1\rangle, |1\rangle^* \}$ and $\{ |2\rangle, |2\rangle^* \}$, respectively, while perturbatively accounting for the remaining basis states using a Schrieffer-Wolff transformation. This gives the effective $2 \times 2$ Hamiltonian of Eq.\ (\ref{eq:valley_transition}).
 
Well in the superconducting phase, for $|\tilde \mu| \gg |\varepsilon_{\theta}|$, the basis kets that we need to project on depend on the signs of $\tilde \mu$ and $\varepsilon_{\theta}$. We choose the phases of the basis kets such, that they can be continuously deformed into the basis vectors used to derive the effective low-energy theory for $\tilde \mu \to \pm \varepsilon_{\theta}$.
Hence, we project onto the basis kets $\{\tfrac{1}{\sqrt{2}}(|1\rangle \pm e^{i \alpha} |2\rangle),\tfrac{1}{\sqrt{2}}(|1\rangle^* \pm e^{-i \alpha} |2\rangle^*)\}$ if $\pm \tilde \mu \gg \pm \varepsilon_{\theta} > 0$ and onto the basis kets $\{\tfrac{1}{\sqrt{2}}(\pm e^{-i \alpha} |1\rangle + |2\rangle), \tfrac{1}{\sqrt{2}}(\pm e^{i \alpha} |1\rangle^* + |2\rangle^*)\}$ if $\pm \tilde \mu \gg \mp \varepsilon_{\theta} > 0$. We then arrive at the effective BdG Hamiltonian of Eq.\ (\ref{eq:valley_singlet}) of the main text.

\section{Chiral p-wave superconductors near \texorpdfstring{$\nu = 0$, $1$}{}}

In the coplanar limit $\theta=0$ the lowest-lying flat band at $E = -t$ touches the quadratic band bottom of the second band of the normal-state Bloch Hamiltonian $H'(\vq)$ of Eq.\ (\ref{eq:H_AFM}) at the $\Gamma$ point $\vq = 0$. The degenerate eigenkets of $H'(\bm{q})$ with energy $-t$ for $\theta = 0$ are denoted $\ket{\pm}$, with
    \begin{align}
      \ket{+} = \frac{1}{\sqrt{3}} \begin{pmatrix} e^{-i 2\pi \chi/3} \\ e^{+i 2\pi \chi/3} \\ 1\end{pmatrix},\quad
      \ket{-} = \frac{1}{\sqrt{3}} \begin{pmatrix} 1 \\ 1 \\ 1 \end{pmatrix}.
      \label{eq:Gamma_eigenkets}
    \end{align}
We can obtain an effective low-energy Hamiltonian upon restricting the BdG Hamiltonian $\mathcal{H}'(\bm{q})$ to the states spanned by the (ordered basis) $\{ \ket{+}, \ket{-}, \ket{+}^*, \ket{-}^*\}$ for $\chi=+1$ and $\{ \ket{-}, \ket{+}, \ket{-}^*, \ket{+}^*\}$ for $\chi=-1$, expanding the diagonal blocks to quadratic order in $\bm{q}$ and linear order in $\theta$, while setting $\bm{q}$ to zero in the off-diagonal blocks. This gives the an effective low-energy theory of the form 
\begin{equation}
  \label{eq:HBdG01}
   \mathcal{H}'(\bm{q}) = \begin{pmatrix} 
   \tilde H({\vq}) - \tilde{\mu}& \tilde \Delta \\
      \tilde \Delta^{\dagger} & -\tilde H (-{\vq})^* + \tilde{\mu} 
    \end{pmatrix},
\end{equation}
where $\mu = -t + \tilde{\mu}$ and the blocks take the form
    \begin{align}
        \label{eq:eff_Gamma_H}
        \tilde H({\vq}) &= \begin{pmatrix} 
            2 \varepsilon_{\theta} + \frac{t}{4} q^2  & 
            -\frac{t}{4} q^2 e^{-2i\alpha} \\
            -\frac{t}{4} q^2 e^{2i\alpha} &  - 2 \varepsilon_{\theta} + \frac{t}{4} q^2
        \end{pmatrix},  \\
        \label{eq:eff_Gamma_Delta}
        \tilde\Delta &= \Delta_{\rm eff}' \begin{pmatrix} 
            i q e^{i \alpha} & i \sqrt{3}  \\
            -i \sqrt{3} & i q e^{-i \alpha}
        \end{pmatrix},
    \end{align}
where we abbreviated $\varepsilon_{\theta} = (3/2) t \chi \theta$.
The band degeneracy of $\tilde{H}(\vq)$ for $\theta = 0$  and $\vq = 0$ is protected by $\mathcal{M}_x = \sigma_x$ and $\mathcal{T} = \sigma_x K$ where $K$ is complex-conjugation and $\sigma_x$ is a Pauli matrix acting in the $2\times2$ space of Eq.\ (\ref{eq:eff_Gamma_H}). Canting out-of-plane ($\theta\neq0$) breaks $\mathcal{T}$ and $\mathcal{M}_x$ (but not $\mathcal{M}_x\mathcal{T}$), lifting the degeneracy. 

The eigenvalues of $\tilde H(\tilde q)$ are 
\begin{equation}
  E_{\pm}(q) = \frac{t}{4} q^2 \pm \sqrt{4 \varepsilon_{\theta}^2 + \tfrac{t^2}{16} q^2}.
\end{equation}
For $\tilde \mu < - 2 |\varepsilon_{\theta}|$ the system is a trivial insulator. For $0 < \tilde \mu < 2 |\varepsilon_{\theta}|$ it is a Chern insulator with Chern number ${\cal C} = -1$ for $\varepsilon_{\theta} > 0$ and ${\cal C} = 1$ for $\varepsilon_{\theta} < 0$ \cite{ohgushi2000spin}. (This corresponds to ${\cal C}_{\rm BdG} = -2$ and $2$, respectively.) The low-energy Hamiltonian (\ref{eq:eff_Gamma_H}) is consistent with the change $\delta {\cal C} = 2$ ({\em i.e.}, $\delta {\cal C}_{\rm BdG} = 4$) upon going through the topological phase transition from $\chi \theta > 0$ to $\chi \theta < 0$. 

If the canting angle $\theta \neq 0$, gapped superconducting phases appear for $-2 |\varepsilon_{\theta}| < \tilde \mu < 0$ and $\tilde \mu > 2 |\varepsilon_{\theta}|$. We first construct effective theories for $\tilde \mu$ just above $-2 |\varepsilon_{\theta}|$ and $2 |\varepsilon_{\theta}|$, {\em i.e.}, for filling fractions just above $\nu = 0$ and $\nu = 1$. We write $\tilde \mu = \varepsilon_{\theta} + d\mu$, abbreviate $s_{\mu} = \mbox{sign}\, \tilde \mu$ and $s_{\theta} = \mbox{sign}\, \varepsilon_{\theta}$, and label the basis states in the $2 \times 2$ matrix notation of Eqs.\ (\ref{eq:eff_Gamma_H}) and (\ref{eq:eff_Gamma_Delta}) as $\ket{1}$ and $\ket{2}$. Restricting the BdG Hamiltonian (\ref{eq:HBdG01}) to the states spanned by $\{\ket{1},\ket{1}^*\}$ for $s_{\mu} s_{\theta} = 1$ and to the states spanned by$\{\ket{2},\ket{2}^*\}$ for $s_{\mu} s_{\theta} = -1$, we find the effective $2 \times 2$ BdG Hamiltonian
\begin{equation}
  \label{eq:Heff01}
  \tilde {\cal H}_{\rm eff}(\vq) = 
  \begin{pmatrix} \tfrac{t}{4} q^2 - d\mu & i q e^{i s_{\mu} s_{\theta} \alpha} \\
  -i q e^{i s_{\mu} s_{\theta} \alpha} & d\mu - \tfrac{t}{4} q^2 \end{pmatrix}.
\end{equation}
For $d\mu > 0$, this Hamiltonian describes chiral p-wave superconductivity, where the superconducting order parameter is of $(q_x - i s_{\theta} q_y)$-type for filling fractions just above $\nu = 0$ and of $(q_x + i s_{\theta} q_y)$-type for filling fractions just above $\nu = 1$.
The phase transition at $d \mu = 0$ is a topological phase transition, at which the Chern number ${\cal C}_{\rm BdG}$ changes by $\mp s_{\theta}$ upon increasing $d\mu$.

\section{Chiral f-wave superconductor near \texorpdfstring{$\nu = 3$}{}}

We now describe the topological superconducting phases for filling fractions just below $\nu=3$, corresponding to $\mu$ just below $2 t$. At these fillings there is a single Fermi pocket around the $\Gamma$ point $\vq = 0$, where the Bloch Hamiltonian (\ref{eq:HM}) has a quadratic dispersion. In the coplanar limit $\theta=0$ the eigenket at $\Gamma$ with energy $2t$ is
\begin{equation}
  \ket{0} = \frac{1}{\sqrt{3}} \begin{pmatrix} e^{+i 2\pi \chi/3} \\ e^{-i 2\pi \chi/3} \\ 1\end{pmatrix}.
  \label{eq:Gamma_lone_eigenket}
\end{equation}
To obtain an effective low-energy Hamiltonian we include the leading-order corrections to the eigenket $\ket{0}$ for small $q$ and $\theta$,
\begin{align}
  \ket{0'} &=
  \ket{0} + \frac{q^2}{24}  \left( e^{+2i\chi\alpha + 2 \theta}\ket{+} + e^{-2i\chi\alpha - 2 \theta}\ket{-} \right),
\end{align}
where the kets $\ket{\pm}$ were defined in Eq.\ (\ref{eq:Gamma_eigenkets}) and we omitted terms of order $q^4$. We then restrict the BdG Hamiltonian $\mathcal{H}'(\bm{q})$ to the states spanned by the (ordered basis) $\{ \ket{0'}, \ket{0'}^*\}$, expanding the diagonal blocks to linear order in $\theta$ and $\Delta_{\rm eff}'$. This gives an effective low-energy theory of the form
\begin{equation}
  \label{eq:Heff3}
  \tilde H(\vq) = \begin{pmatrix} -\tfrac{t}{2} q^2 - \tilde \mu &
  \tilde \Delta(\vq) \\ \tilde \Delta(\vq)^* & \tfrac{t}{2} q^2 + \tilde \mu
  \end{pmatrix},
\end{equation}
with $\mu = 2 t + \tilde \mu$ and 
\begin{equation}
  \tilde \Delta = - \frac{i}{6} \Delta_{\rm eff}' q^3
    (\cos 3 \alpha - i \chi \theta \sin 3 \alpha).
\end{equation}
In the coplanar limit, this describes a soft-gap superconductor with pairing $\sim\Delta_{\rm eff}' q^3 \cos(3\alpha)$ and nodes for 6 momenta along the Fermi surface. Canting the spin texture opens a gap, giving rise to a chiral f-wave superconductor with a chirality set by that of the spin texture and the canting angle. 

\section{Phase diagram}

The Chern numbers ${\cal C}_{\rm BdG}$ in the phase diagram of Fig.\ \ref{fig:Panel}c have been calculated independently using three methods: (i) direct calculation from the BdG Hamiltonian ${\cal H}'(\vq)$ of Eq. \ (\ref{eq:HM}) using the Wilson-loop method \cite{alexandradinata2014}, (ii) by counting the number of chiral boundary states in a strip geometry, and (iii) by tracking the change of ${\cal C}_{\rm BdG}$ across topological phase boundaries using the effective low-energy descriptions near integer filling fractions $\nu$. Here, we  provide additional details for the methods (ii) and (iii).

\begin{figure*}[t]
	\includegraphics[width=1\textwidth]{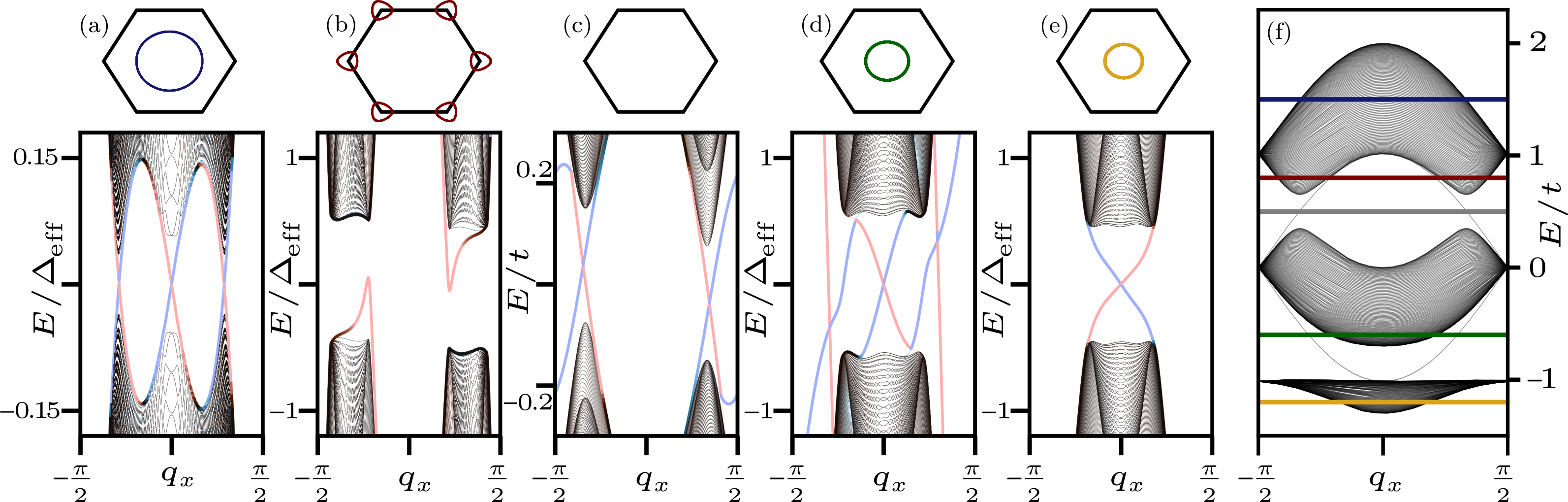}
	\caption{(a)-(e) Fermi surfaces (top) and spectra of the BdG Hamiltonian (\ref{eq:HM}) (bottom) for a strip geometry along the $x$ direction for five representative values of the chemical potential $\mu$, as indicated in the rightmost panel (f). States localized near the lower boundary of the strip at $y=0$ are colored in red and those near the upper boundary are colored in blue. From left to right, the values of the chemical potential in panels (a)-(e) are $\mu = -1.2 t$, $-0.6 t$, $0.5 t$, $0.8 t$, and $1.5 t$. The canting angle is $\theta = 0.1$ and $\Delta_{\rm eff} = 0.05 t$, except for the spectra for $\mu = 1.5 t$, where we set $\Delta_{\rm eff} = 0.2 t$.}
	\label{fig:Boundary}
\end{figure*}

(ii): Figure \ref{fig:Boundary} shows the spectrum of the BdG Hamiltonian (\ref{eq:HM}) for a strip in the $x$ direction, with $0 \le y \le W$, for five representative values of the chemical potential $\mu$, as indicated in the figure. The Chern number ${\cal C}_{\rm BdG}$ is found as the difference of the numbers of right-moving and left-moving boundary states localized near the boundary at $y=0$ (shown in red).

(iii): For the insulating phases, shown in white in the phase diagram of Fig.\ \ref{fig:Panel}c, the Chern number $C_{\rm BdG}$ equals twice the Chern number ${\cal C}$ of the normal-state insulator, because of the doubling of the Hilbert space in the BdG formalism. To find the Chern numbers of the superconducting phases, shown in red in Fig.\ \ref{fig:Panel}c, we track the change of ${\cal C}_{\rm BdG}$ across the topological phase transition to the adjacent insulating phase. This can be done using the effective low-energy theories for filling fractions above $\nu = 0$, above $\nu = 1$, below and above $\nu = 2$, and below $\nu = 3$. (Note that the effective low-energy theories cannot be used to directly calculate ${\cal C}_{\rm BdG}$, because this calculation involves an integral of the Berry curvature over the entire Brillouin zone. The effective low-energy theories can, however, be reliably used for a calculation of the change of ${\cal C}_{\rm BdG}$, since the change of the Berry curvature is confined to the immediate vicinity of the gap closing points at the topological phase transition.) Specifically: 
\begin{itemize}
\item Upon increasing $d\mu$, the effective low-energy theory of Eq.\ (\ref{eq:Heff01}) gives a change $\delta {\cal C}_{\rm BdG} = s_{\theta} = \mbox{sign}(\theta \chi)$ for filling fraction near $\nu =1$ and $\delta {\cal C}_{\rm BdG} = - s_{\theta}$ for filling fraction near $\nu = 2$; 
\item Starting from the insulating phase at $\nu = 2$, the effective low-energy theory of Eq.\ (\ref{eq:valley_transition}) gives a change $\delta {\cal C}_{\rm BdG} = 2 s_{\theta}$ upon decreasing or increasing the chemical potential; 
\item Upon decreasing $\tilde \mu$, the effective low-energy theory of Eq.\ (\ref{eq:Heff3}) gives a change $\delta {\cal C}_{\rm BdG} = -3 s_{\theta}$ for filling fraction near $\nu = 3$. 
\end{itemize}
These changes correspond precisely to the changes shown in Fig.\ \ref{fig:Panel}c. 

\section{Josephson junction}

For the numerical calculation of the supercurrent in Fig.\ \ref{fig:Panel}f, we use the BdG Hamiltonian (\ref{eq:HM}) of the fully-polarized limit. We consider a chiral antiferromagnet of length $L'$ (in the $y$ direction) and width $W$ (in the $x$ direction). We apply periodic boundary conditions in the $x$ direction and twisted boundary conditions with an additional flux $\varphi$ in the $y$ direction. We choose $L' = 400 \sqrt{3}$ to ensure that the ring is sufficiently large to avoid finite size effects. As we consider the short-junction limit, we do not need to explicitly insert a strip without proximity superconductivity. We obtain the (positive) eigenvalues $\varepsilon_i(k_x,\varphi)$ of the BdG Hamiltonian at fixed $k_x$ and flux $\varphi$ and numerically calculate the total energy (up to $\varphi$-independent terms) as
\begin{equation}
  E(\varphi) = - \sum_{i} \sum_{k_x} \varepsilon_i(k_x,\varphi).
\end{equation}
The supercurrent is then obtained from \cite{beenakker1991d} \begin{equation}
  I(\varphi) = \frac{e}{\hbar} \frac{\partial E(\varphi)}{\partial \varphi}.
\end{equation}
The results of this calculation corresponds to the red and blue curves in Fig.\ \ref{fig:Panel}f, showing good agreement with the analytical prediction (see main text for a discussion of the small discrepancies).

\section{Kagome layer uniformly coupled to a superconductor}

In the main text, we consider a model in which the tunnel coupling of a single magnetic kagome layer to a conventional spin-singlet supercondutor effectively induces superconductivity on the bonds of the upward-facing triangle in the kagome lattice, but not on the bonds of the downward-facing triangle. This way, the coupling to the superconductor breaks the inversion symmetry of the kagome lattice, which is a prerequisite for inducing spin-polarized superconductivity.

Here we  consider a model with spatially uniform coupling to a superconductor, so that the proximity-induced superconductivity does not break the inversion symmetry in the kagome layer. In a lattice model, one may obtain a spatially uniform coupling to the superconductor by placing the superconductor on the dual of the kagome lattice. (The dual of the kagome lattice is a lattice of sites located on the bonds of the kagome lattice.) This way, superconductivity is imposed equally on all bonds of the kagome lattice, not just on the bonds of the upward-facing triangles.

The BdG Hamiltonian that describes proximity-induced superconductivity in this system is Eq.\ (\ref{eq:HBDG}) with
\begin{equation}
  \label{eq:DeltaUniform}
  \Delta(\vq) = 2 \Delta_{\rm eff} \sigma_y 
  \begin{pmatrix} 2 & \cos q_3 & \cos q_2 \\ \cos q_3 & 2 & \cos q_1 \\
  \cos q_2 & \cos q_1 & 2 \end{pmatrix}.
\end{equation}
Here $\Delta_{\rm eff} = |\tau|^2/\Delta_0^*$, with $\Delta_0$ the order parameter of the superconductor and $\tau$ the tunneling matrix element for hopping between neighboring sites of the magnetic and superconducting layers. Proceeding as in the main text, we find a Hamiltonian of the form (\ref{eq:HM}) for the fully-polarized limit, with
\begin{align}
  \Delta'(\vq) =&\, 2 \Delta_{\rm eff}'  \\ \nonumber &\, \mbox{} \times
  \begin{pmatrix} 0 & \cos q_3 & -e^{i \omega} \cos q_2 \\ -\cos q_3 & 0 &  e^{-i \omega} \cos q_1 \\
   e^{i \omega} \cos q_2 & - e^{-i \omega} \cos q_1 & 0 \end{pmatrix},
\end{align}
with $\Delta_{\rm eff}'$ given below Eq.\ (\ref{eq:Delta_eff_prime}).

Inversion symmetry imposes that $\Delta'(\vq)$ is an antisymmetric matrix, whereas the normal-state Hamiltonian $H'(\vq) = H'(-\vq)$. As a result, matrix elements of $\Delta'(\vq)$ between an eigenstate of $H'(\vq)$ and its particle-hole conjugate, i.e., an eigenstate of $H'(-\vq)^*$, must vanish. This rules out a superconducting gap in the half-metallic limit to leading (first) order in $\Delta_{\rm eff}$ for non-degenerate bands.

\section{Proximity-induced superconductivity in a magnetic bilayer}
\label{app:bilayer}

In the fully-polarized limit, there is no superconducting gap to first order in $\Delta_{\rm eff}$ if the coupling to the superconductor is spatially uniform and if the normal-state Hamiltonian has in-plane inversion symmetry. In-plane inversion symmetry is broken in a bilayer magnet if the two magnetic kagome layers are shifted in-plane as in Mn$_3$Ge, so that the downward-facing triangles of one layer are on top of the upward-facing triangles of the other layer. We here show that such a magnetic bilayer admits spin-polarized superconductivity if only one of the two layers is proximity-coupled to a superconductor in a spatially uniform manner. (The coupling of only one magnetic layer to the superconductor ensures that there is no inversion symmetry that involves interchanging the two layers, too.) 

\begin{figure}[t]
\includegraphics[width=0.75\columnwidth]{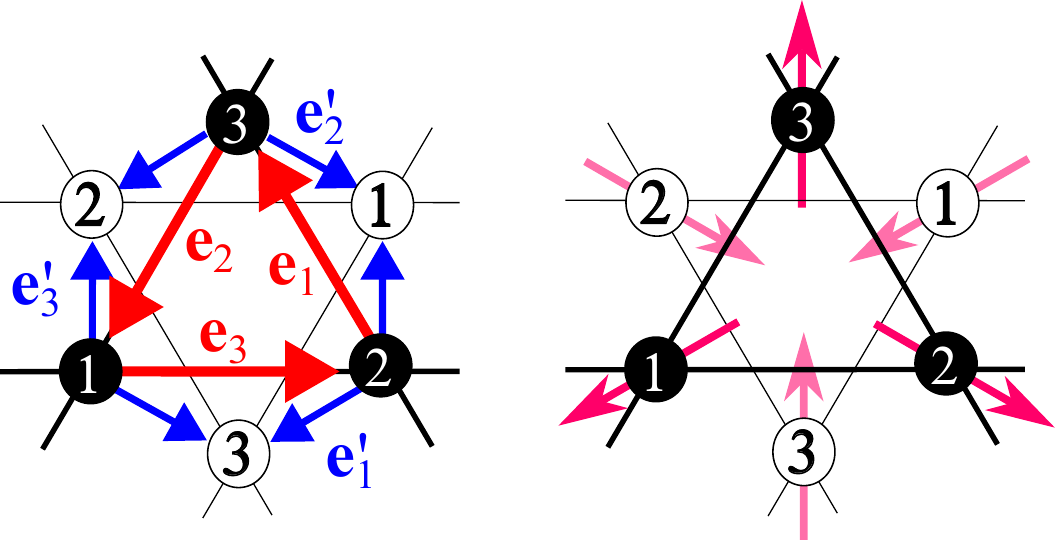}
\caption{Labeling convention (left) and orientation of the exchange field (right) for the unit cell of the kagome bilayer.}
\label{fig:unitcell_bilayer}
\end{figure}

The BdG Hamiltonian of such a bilayer kagome antiferromagnet coupled to a spin-singlet superconductor is of the form (\ref{eq:HBDG}), where each block now has size $12 \times 12$. The normal-state Bloch Hamiltonian of the itinerant magnetic bilayer reads 
\begin{equation}
  H_{i\alpha,j\beta}(\vq) = -\vsigma \cdot \vh_j \delta_{ij} \delta_{\alpha\beta} + \sigma_0 T_{i\alpha,j\beta}(\vq),
  \label{eq:H_Bloch_bilayer}
\end{equation}
where $i,j = 1,2,3$ refer to the three sites in the unit cell for each layer, see Fig.\ \ref{fig:unitcell_bilayer}, and $\alpha,\beta=1,2$ is the layer index. The exchange field is given by Eq.\ (\ref{eq:hh}) and the intra-layer hopping $T_{i\alpha,j\alpha}(\vq) = T_{ij}(\vq)$, $\alpha=1,2$, with $T_{ij}(\vq)$ given in Eq.\ (\ref{eq:Hspinful}). The inter-layer hopping reads
\begin{equation}
  T_{1,2}(\vq) = T_{2,1}(\vq)^{*} = - \tau \begin{pmatrix}
  0 & e^{i q_3'} & e^{i q_2'} \\
  e^{i q_3'} & 0 & e^{i q_1'} \\
  e^{i q_2'} & e^{i q_1'} & 0 \end{pmatrix},
\end{equation}
where $\tau$ is the inter-layer hopping amplitude and $q_i' = \vq \cdot \ve_i'$, where $\ve_i' = (1/\sqrt{3}) \ve_z \times \ve_i$, $i=1,2,3$, see Fig.\ \ref{fig:unitcell_bilayer}. (We take the kagome layers in the $xy$ plane.) For the explicit choice of Eq.\ (\ref{eq:exy}), one has 
\begin{align}
  \ve_1' =&\, - \tfrac{1}{2} \ve_x - \tfrac{1}{6} \ve_y \sqrt{3}, \nonumber \\
  \ve_2' =&\, \tfrac{1}{2} \ve_x - \tfrac{1}{6} \ve_y \sqrt{3}, \nonumber \\
  \ve_3' =&\, \tfrac{1}{3} \ve_y \sqrt{3}.
\end{align}
For the proximity-induced superconducting order parameter we take (cf.\ Eq.\ (\ref{eq:DeltaUniform}))
\begin{equation}
  \Delta_{\alpha,\beta}(\vq) = 2 \Delta_{\rm eff} \sigma_y \delta_{\alpha,1} \delta_{\beta,1}
  \begin{pmatrix} 2 & \cos q_3 & \cos q_2 \\ \cos q_3 & 2 & \cos q_1 \\
  \cos q_2 & \cos q_1 & 2 \end{pmatrix}.
\end{equation}

We again take the fully-polarized limit and arrive at a BdG Hamiltonian similar to Eq.\ (\ref{eq:HM}), but with $6 \times 6$ blocks $H'_{\alpha,\beta}(\vq)$, with the diagonal blocks $H'_{\alpha,\alpha}(\vq) = H'(\vq)$, with $H'(\vq)$ given by Eq.\ (\ref{eq:H_AFM}), and
\begin{equation}
  H'(\vq)_{1,2} = - \begin{pmatrix}
  0 & \tau' e^{i q_3'} & \tau'^* e^{i q_2'} \\
  \tau'^* e^{i q_3'} & 0 & \tau' e^{i q_1'} \\
  \tau' e^{i q_2'} & \tau'^* e^{i q_1'} & 0 \end{pmatrix},
\end{equation}
where
\begin{equation}
  \tau' = \tau e^{i \omega/2}[\cos(\omega/2) - i \sin(\omega/2) \sin \theta].
\end{equation}
The effective superconducting order parameter in the fully-polarized limit reads
\begin{align}
  \Delta'_{\alpha,\beta}(\vq) =&\, 2 \Delta_{\rm eff}'  \delta_{\alpha,1} \delta_{\beta,1} \\ \nonumber &\, \mbox{} \times
  \begin{pmatrix} 0 & \cos q_3 & -e^{i \omega} \cos q_2 \\ -\cos q_3 & 0 &  e^{-i \omega} \cos q_1 \\
   e^{i \omega} \cos q_2 & - e^{-i \omega} \cos q_1 & 0 \end{pmatrix},
\end{align}
with $\Delta_{\rm eff}' = \Delta_{\rm eff} e^{-i \phi} \cos \theta \sin \omega$.

We now obtain the effective low-energy theory near filling fraction $\nu = 4$, where the fully-polarized bilayer has Fermi pockets around the $K$ and $K'$ points. For $\vq$ in the vicinity of the $K$ point, we have (see also Eq.\ (\ref{eq:qKK}))
\begin{align} 
q_1' =&\, -\tfrac{\pi}{3} - \tfrac{1}{2} \tilde q_x  - \tfrac{1}{6} \tilde q_y \sqrt{3}, \nonumber \\  q_2' =&\, \tfrac{\pi}{3} + \tfrac{1}{2} \tilde q_x - \tfrac{1}{6} \tilde q_y \sqrt{3}, \nonumber \\ q_3' =&\, \tfrac{1}{3} \tilde q_y \sqrt{3}.
\end{align}
We project onto the (ordered) basis $\{ |\pm\rangle_1,|\mp\rangle_1, |\pm\rangle_2,|\mp\rangle_2,|\pm\rangle^*_2,|\mp\rangle^*_2,|\pm\rangle^*_1,|\mp\rangle^*_1 \}$ if $\chi = \pm 1$, see Eq.\ (\ref{eq:Keigenkets}), expand the intra-layer terms to first order in $\tilde{\vq}$, and set $\tilde{\vq}$ to zero for all other terms. This gives a Hamiltonian for electron-like excitations near $K$ and hole-like excitations near $K'$ of the form (\ref{eq:HK}), but with $4 \times 4$ blocks $\tilde H_+(\tilde q)$, $\tilde H_-(-\tilde q)^*$, and $\tilde \Delta$,
\begin{align}
  \label{eq:HK_bilayer}
  \tilde H_{\pm}(\tilde \vq) =&\,
  \begin{pmatrix} \varepsilon_{\theta} & \mp v \tilde q  e^{-i \alpha} & 0 & \tau \\
    \mp v \tilde q  e^{i \alpha} & -\varepsilon_{\theta} & 0 & 0 \\
    0 & 0 & \varepsilon_{\theta} & \mp v \tilde q  e^{-i \alpha} \\
    \tau & 0 & \mp v \tilde q  e^{i \alpha} & -\varepsilon_{\theta} \end{pmatrix}, \\
    \label{eq:DeltaK_bilayer}
  \tilde \Delta =&\,
  \begin{pmatrix} 0 & 0 & 0 & i \Delta_{\rm eff}' \sqrt{3} \\
    0 & 0 & -i \Delta_{\rm eff}' \sqrt{3} & 0 \\
    0 & 0 & 0 & 0 \\
    0 & 0 & 0 & 0 \end{pmatrix},
\end{align}
where we again abbreviated $\varepsilon_{\theta} = (3/2) t \theta \chi$, $v = (1/2) t \sqrt{3}$, and $\tilde q e^{i \alpha} = \tilde q_x + i \tilde q_y$. The full BdG Hamiltonian also contains a block that describes electron-like excitations near $K'$ and hole-like excitations near $K$, which is the particle-hole conjugate of Eqs.\ (\ref{eq:HK_bilayer}) and (\ref{eq:DeltaK_bilayer}).

At the $K$ and $K'$ points, the states deriving from the first and fourth row/column in Eq.\ (\ref{eq:HK_bilayer}) are gapped out by the interlayer coupling, so that for $|\tilde \mu| \ll \tau$ we may further project onto states spanned by the second and third columns/rows using a Schrieffer-Wolff transformation. This again gives an effective $4 \times 4$ BdG Hamiltonian of the form (\ref{eq:HK}), but now with $2 \times 2$ blocks, which read
\begin{align}
  \tilde H_{\pm}(\tilde \vq) =&\,
  \begin{pmatrix}
    \varepsilon_{\theta} & - D \tilde q^{2}  e^{-2 i \alpha} \\
    - D \tilde q^{2}  e^{2 i \alpha} & - \varepsilon_{\theta}
  \end{pmatrix}, \label{eq:H2BdG1} \\ \label{eq:H2BdG2}
  \tilde \Delta =&\,
  \begin{pmatrix} 0 & 0 \\ -i \Delta_{\rm eff}' \sqrt{3} & 0 \end{pmatrix},
\end{align}
where we further abbreviated $D = v^2/\tau = 3t^2/4\tau$. 

To find the effective theory for $|\varepsilon_{\theta}| \ll |\tilde \mu| \ll \tau$, we again choose the phase of the basis vectors such that they continuously evolve into the appropriate basis vectors for $\tilde \mu \to \pm \varepsilon_{\theta}$. We then find the effective $2 \times 2$ BdG Hamiltonian
\begin{align}
  {\cal H}_{\rm eff}(\tilde \vq) =&\,
  \begin{pmatrix}
    \pm D (\tilde q^2 - q_{\rm F}^2) &
    \pm \tfrac{i}{2} \Delta_{\rm eff}' \sqrt{3} e^{\mp 2 i s_{\theta} \alpha} \\
    \mp \tfrac{i}{2} \Delta_{\rm eff}' \sqrt{3} e^{\pm 2 i s_{\theta} \alpha} & \mp D (\tilde q^2 - q_{\rm F}^2)
  \end{pmatrix},
\end{align}
where $\tilde \mu = \pm D q_{\rm F}^2$ and $s_{\theta} = \mbox{sign}\, (\varepsilon_{\theta})$. This effective low-energy Hamiltonian describes a fully-gapped valley-singlet superconductor, consistent with the numerically obtained spectrum in Fig.\ \ref{fig:Bilayer} of the main text.

\bibliography{refs.bib}

\end{document}